\documentclass[11pt]{article}
\usepackage{latexsym}
\def\thefootnote{\fnsymbol{footnote}}

\def\tk{{\tilde k}}

\def\ts{{\tilde s}}
\def\tL{{\tilde L}}

\def\hE{{\hat E}}

\def\tF{{\tilde F}}

\def\hR{{\hat R}}

\def\bl{\bar{\lambda}}

\def\hL{\hat{L}}

\def\hM{\hat{M}}
\def\hS{\hat{S}}

\def\hK{\hat{K}}

\def\hb{\hat{b}}

\def\dx{\delta_X}

\def\r{\right|}
\def\l{\left.}
\def\[{\left [}
\def\]{\right ]}
\def\({\left (}
\def\){\right )}

\def\pp{\partial}

\def\R{\bar{R}}
\def\br{\bar{r}}

\def\ph{\bar{\phi}}

\def\bth{\bar{\theta}}
\def\bTh{\bar{\Theta}}
\def\ux{$U(1)_X$}

\def\vev{$vev$}

\def\H{\bar{H}}
\def\h{\bar{h}}

\def\T{\bar{T}}

\def\s{\bar{s}}

\def\M{\bar{M}}

\def\hbM{\hat{\M}}

\def\re{{\rm Re}}
\newcommand{\beq}{\begin{equation}}
\newcommand{\eeq}{\end{equation}}
\newcommand{\bea}{\begin{eqnarray}}
\newcommand{\eea}{\end{eqnarray}}
\def\bF{\bar{F}}

\def\f{\bar{f}}

\def\L{{\cal L}}
\newcommand{\superint}{\int \diff^{4}\theta \, }

\newcommand{\diff}{\mbox{d}}

\def\D{{\cal D}}
\def\bD{{\bar{\cal D}}}

\newcommand{\WaWa}{ W^{\alpha}{ W}_{\alpha}}

\newcommand{\DaDa}{{\cal D}^2}
\newcommand{\DbDb}{{\bar{\cal D}}^2}
\newcommand{\Da}{{\cal D}^{\alpha}}

\newcommand{\Dc}{{\cal D}_{\alpha}}

\def\vx{V_X}

\def\pp{\partial}
\def\rvev{\right\rangle}
\def\lvev{\left\langle}

\def\hel{\hat\ell}

\def\del{\delta}

\newcommand{\bbar}[1]{{\overline{#1}}}

\newcommand{\ord}[1]{{{\cal O}(#1)}}
\newcommand{\myref}[1]{(\ref{#1})}
\newcommand{\beqa}{\begin{eqnarray}}
\newcommand{\eeqa}{\end{eqnarray}}
\newcommand{\nnn}{ \nonumber \\ }
\newcommand{\p}{\partial}
\newcommand{\hc}{{\rm h.c.}}
\newcommand{\W}{{\cal W}}
\newcommand{\Wb}{{\bbar{{\cal W}}}}
\newcommand{\chiproj}{(\bD^2 - 8R)}
\newcommand{\hchiproj}{(\hat \bD^2 - 8 \hat R)}
\newcommand{\bchiproj}{(\D^2 - 8 \bar R)}
\newcommand{\Zbf}{{{\bf Z}}}
\newcommand{\fourth}{{1 \over 4}}
\newcommand{\uone}{$U(1)$}
\newcommand{\Phibar}{\bar \Phi}
\newcommand{\evev}[1]{{\langle #1 \rangle}}
\newcommand{\bigvev}[1]{{\left\langle #1 \right\rangle}}
\newcommand{\Y}{{\cal Y}}
\newcommand{\Dk}{\Delta k}
\newcommand{\third}{{1 \over 3}}
\newcommand{\tr}{\mathop{{\hbox{tr} \, }}\nolimits}
\newcommand{\alphadot}{{\dot \alpha}}

\newcommand{\half}{{1 \over 2}}
\newcommand{\Dau}{\hat \D^{\alpha}}
\newcommand{\Dad}{\hat \D_{\alpha}}
\newcommand{\Dbu}{\hat \D^{\beta}}
\newcommand{\Dbd}{\hat \D_{\beta}}

\begin{document}

\begin{titlepage} 
            
\hfill   LBNL-50097
            
\hfill   UCB-PTH-02/14
            
\hfill   hep-th/0204100

\hfill   April 2002

\begin{center}

\vspace{18pt}
{\bf \Large Modular Invariant Anomalous ${\bf U(1)}$ Breaking}

\vspace{18pt}
 Mary K. Gaillard\footnote{E-Mail: {\tt MKGaillard@lbl.gov}}
{\em and} Joel Giedt\footnote{E-Mail: {\tt JTGiedt@lbl.gov}}

\vspace{18pt}

{\em Department of Physics, University of California, \\
and Theoretical Physics Group, 50A-5101, \\
Lawrence Berkeley National Laboratory, Berkeley, 
CA 94720 USA.}\footnote{This work was supported in part by the
Director, Office of Science, Office of High Energy and Nuclear
Physics, Division of High Energy Physics of the U.S. Department of
Energy under Contract DE-AC03-76SF00098 and in part by the National
Science Foundation under grant PHY-0098840.}\\[.1in]

\vspace{18pt}

\end{center}

\begin{abstract}
We describe the effective supergravity theory present
below the scale of spontaneous gauge symmetry breaking
due to an anomalous $U(1)$, obtained by integrating
out tree-level interactions of
massive modes.  A simple case is examined in some
detail.  We find that the effective theory can be
expressed in the linear multiplet formulation,
with some interesting consequences.  Among them,
the modified linearity conditions lead to new interactions
not present in the theory without an anomalous $U(1)$.
These additional interactions are compactly
expressed through a superfield functional.
\end{abstract}

\end{titlepage}

\newpage
\renewcommand{\thepage}{\roman{page}}
\setcounter{page}{2}
\mbox{ }

\vskip 1in

\begin{center}
{\bf Disclaimer}
\end{center}

\vskip .2in

\begin{scriptsize}
\begin{quotation}
This document was prepared as an account of work sponsored by the United
States Government. Neither the United States Government nor any agency
thereof, nor The Regents of the University of California, nor any of their
employees, makes any warranty, express or implied, or assumes any legal
liability or responsibility for the accuracy, completeness, or usefulness
of any information, apparatus, product, or process disclosed, or represents
that its use would not infringe privately owned rights. Reference herein
to any specific commercial products process, or service by its trade name,
trademark, manufacturer, or otherwise, does not necessarily constitute or
imply its endorsement, recommendation, or favoring by the United States
Government or any agency thereof, or The Regents of the University of
California. The views and opinions of authors expressed herein do not
necessarily state or reflect those of the United States Government or any
agency thereof of The Regents of the University of California and shall
not be used for advertising or product endorsement purposes.
\end{quotation}
\end{scriptsize}

\vskip 2in

\begin{center}
\begin{small}
{\it Lawrence Berkeley Laboratory is an equal opportunity employer.}
\end{small}
\end{center}

\newpage
\renewcommand{\thepage}{\arabic{page}}
\setcounter{page}{1}
\def\thefootnote{\arabic{footnote}}
\setcounter{footnote}{0}

\noindent
The existence of anomalous \uone\ factors 
(hereafter denoted \ux)
in the effective theories derived from superstrings
is generic.  Indeed, in a recent study \cite{Gie01c}
of a large
class of standard-like heterotic $Z_3$ orbifold
models, it was found that 168 of 175 models
had an anomalous \ux.  Since the underlying
theory is anomaly free, it is known \cite{UXR}
that the apparent anomaly is canceled by 
a four-dimensional version of the
Green-Schwarz (GS) mechanism \cite{GS84}.  This
leads to a Fayet-Illiopoulos (FI) term in
the effective supergravity Lagrangian.  Ignoring
nonperturbative corrections to the
dilaton K\"ahler potential,
\beq
D_X = \sum_i {\p K \over \p \phi^A} q^X_A \phi^A + \xi,
\qquad \xi = {g_s^2 \tr Q_X \over 192 \pi^2} m_P^2,
\label{eq1}
\eeq
where $K$ is the K\"ahler potential, $q_A^X$ is the
\ux\ charge of the (complex) scalar matter field $\phi^A$,
$\xi$ is the FI term, $Q_X$ is the charge generator
of \ux, $g_s$ is the unified (string scale) gauge coupling,
and $m_P=1/\sqrt{8\pi G} = 2.44 \times 10^{18}$ GeV is
the reduced Planck mass.  In the remainder we
work in units where $m_P=1$.

At tree-level in the underlying theory,
the chiral dilaton formulation has $g_s^2 = 1/\re \evev{s}$,
where $s= S|$ is the lowest component of the
chiral dilaton superfield $S$.  However, once
higher order and nonperturbative corrections
are taken into account the chiral dilaton
formulation becomes inconvenient.  The dual
linear multiplet formulation---which relates
a (modified) linear superfield $L$ to $\{ S,\bar S\}$
through a duality transformation---provides
a more convenient arrangement of superfield degrees
of freedom due to the neutrality of $L$
with respect to target-space duality transformations
(hereafter called {\it modular transformations}).
In the limit of vanishing nonperturbative
corrections to the dilaton K\"ahler potential,
$g_s^2 = 2 \evev{\ell}$, where $\ell = L|$.
Throughout this article we use the linear multiplet
formulation \cite{sjg,LMULT}.  Except where noted below, we
use the $U(1)_K$ superspace formalism \cite{BGGM,BGG01}.
(For a review of the $U(1)_K$ superspace
formalism see \cite{BGG01}; for a review of
the linear multiplet formulation see \cite{GG99}.)

In the linear multiplet formulation, the FI term
becomes
\beq
\xi(\ell) = {2 \ell \cdot \tr Q_X \over 192 \pi^2}.
\label{eq2}
\eeq
Consequently, the background dependence of the FI
term in \myref{eq2} arises from $\evev{\ell}=\evev{L|}$.
The FI term induces nonvanishing
vacuum expectation values ({\it vev}'s) for
some scalars $\phi^i$ as the scalar potential
drives $\evev{D_X} \to 0$, if supersymmetry is unbroken.
The nonvanishing \vev's in the
supersymmetric vacuum phase can be related to the
FI term.  Then
$\evev{L|}$ serves as an order parameter for the vacuum
and all nontrivial \vev's can be written as some
fraction of $\evev{L|}$.  {\it Our approach in
what follows will be to promote this to a
superfield redefinition.}

Our starting point is the effective supergravity
model of gaugino condensation developed 
by Bin\'etruy, Gaillard and Wu (BGW) \cite{BGWa,BGWb}
as well as subsequent elaborations by
Gaillard, Nelson and Wu \cite{GNW99,GN00}.
A significant modification is the inclusion
of a \ux\ factor in the gauge group and the
corresponding GS counterterm in the effective
Lagrangian.
The effective Lagrangian at the string scale is defined by
\beq
\L = \superint\tL + \L_{th} + \L_Q .
\label{str}
\eeq
The first term is the superspace integral of the
real superfield functional
\beq
\tL = E\[- 3 + 2Ls(L) + L\(bG - \dx\vx\)\].
\label{oit}
\eeq
This contains the usual kinetic
term $(-3)$ in the $U(1)_K$ formalism, as well as
tree-level terms with explicit dependence on
the (modified) linear superfield $L$.
The contribution $2 L s(L)$ includes 
the gauge kinetic term
of the more conventional supergravity
formulation.  
(In the dual chiral formulation 
$s(L) \to {\rm Re}s$.)
In the BGW articles \cite{BGWa,BGWb}, this
was written in terms of a functional $f(L)$
such that $2Ls(L) = 1 + f(L)$.
Note that $s(\evev{L|}) = g_s^{-2}$ determines
the unified (string scale) gauge coupling $g_s$.
The contribution
$L\(bG - \dx\vx\)$ provides the GS
counter-terms which cure field theoretic
anomalies associated with modular
and \ux\ transformations.  Here, $V_X$
is the \ux\ vector superfield and $G$
is defined by\footnote{
In our considerations
we oversimplify by considering only the
three ``diagonal'' K\"ahler moduli $T^I = T^{II}$
$(I=1,2,3)$, present in each of the
$Z_N$ and $Z_M \times Z_N$ six-dimensional
orbifolds, and transforming under an
$SL(2,\Zbf)^3$ subgroup of the full
modular duality group \cite{FKP86,mdg}.}
\beq
G = \sum_I g^I, \qquad g^I = -\ln(T^I + \T^I).
\label{gId}
\eeq

A prominent advantage of the $U(1)_K$ formalism
is that Weyl rescalings are performed at the
superfield level, and no rescalings are necessary
at the component field level to obtain a canonical
Einstein term.
For example in the Lagrangian \myref{str} which
we start with, we require
\beq
k'(L) + 2Ls'(L) = 0,
\label{canon}
\eeq
where $k(L)$ is the $L$-dependent part of the K\"ahler
potential.  
Of chief concern in what follows will be the
maintenance of the canonical normalization for the Einstein
term---concurrent to field redefinitions.  
Therefore we lay out a general prescription
for determining the necessary {\it Einstein condition}
from $\L$ rewritten in a new field basis.

The relevant part is \myref{oit}.  We define $M$ to
stand collectively for the fields which are to be
regarded as independent of $L$ in a given basis.  We
then define the functional $S$ by the identification
\beq
\tL \equiv E[-3 + 2L S(L,M)].
\label{jut}
\eeq
The Einstein condition holds provided
\beq
\( {\p K \over \p L}\)_M + 2 L \( {\p S \over \p L}\)_M = 0.
\label{aab}
\eeq
Here, the subscripts on parentheses
instruct us to hold constant
under differentiation the fields denoted
collectively by $M$.

It can be seen from \myref{oit} that
\beq
S=s(L) + \half (bG-\dx V_X)
\label{uiw}
\eeq
and that \myref{aab} applied to \myref{uiw}
is equivalent to \myref{canon}.

It is our intent to integrate out the modes which
become heavy due to the FI gauge symmetry
breaking.  Clearly the \ux\ vector multiplet
becomes massive.  Then the most relevant
parts of the Lagrangian are those where
the chiral field strength $\W_X$ appears.
It is important to keep in mind the modified
linearity conditions
\beq
\chiproj L = - \sum_a (\W \W)_a, \qquad
\bchiproj L = - \sum_a (\Wb \Wb)_a.
\label{abc}
\eeq
Because of this, the kinetic and 
$L$-dependent parts of the Lagrangian
are the focus of most of our attention.
Our manipulations involve superfield redefinitions
which are intended to give $\tL$
a form where heavy modes are apparent
and are not linearly coupled to light modes.
Truncation of the field content to
the new light field basis then
accounts for tree level exchange of
heavy modes.

Note that we are not using $U(1)_K$ superspace 
for the anomalous \ux.  
That is, the covariant derivatives 
used to define component fields contain
the connections for the unbroken gauge group $G_C$,
but not \ux.  The vector superfield
$V_X$ has to be introduced {\it explicitly} (as opposed
to the geometric method of $U(1)_K$ superspace)
both to regulate
the QFT loops \cite{PVREG}, and in the GS term \cite{UXR}.
However, there is no problem including a
Chern-Simons superfield for \ux\ in the duality
transformation (discussed below) giving $L$, so the modified
linearity conditions \myref{abc} still lead to
gauge kinetic terms for \ux.

The second term in \myref{str}, $\L_{th}$, accounts for
threshold effects due to heavy states
above the string scale.  The third term, 
$\L_Q$, gives the quantum corrections 
from states below the string scale
to the effective Lagrangian.  
The K\"ahler potential is kept to
leading order\footnote{This approximation may not be
justified for fields $\Phi^A$ with large \vev's; corrections
to it will be considered elsewhere.} in matter fields $\Phi^A$.
For the compactification moduli $T^I$
we keep the well-known terms which
have been extracted from the dimensional reduction of
ten-dimensional supergravity \cite{kpota,FKP86}
or the matching to four-dimensional string amplitudes \cite{kpotb}.
The linear multiplet contribution
is allowed a nonperturbative contribution
$g(L)$ which will be exploited for stabilization.
Altogether,
\beqa
K &=& k(L) + G + \sum_A e^{G^A + 2 q_A^X V_X}|\Phi^A|^2, \nnn
k(L) &=& \ln L + g(L), 
\qquad G^A \equiv \sum_I q^A_I g^I.
\label{kpt}
\eeqa
For each $\Phi^A$, the \ux\ charge is denoted 
$q_A^X$ while $q^A_I$ are the modular weights.  
The convention chosen in
\myref{kpt} implies \ux\ gauge invariance
corresponds to the transformation
\beq 
\vx \to  \vx' = \vx + {1\over2}\(\Theta + \bTh\), 
\quad \Phi^A \to \Phi'^A = e^{-q^A_X\Theta}\Phi^A.
\eeq

The threshold contribution from heavy
string excitations is\footnote{A slight change
in conventions has been made here and
in $\L_{Q}$ below, versus
Refs.~\cite{BGWb,gt,GNW99}, 
involving factors of $8 \pi^2$.}
\beq
\L_{th} = -\superint\,{E\over 8R}
\sum_{a,I} b_a^I (\W \W)_a
\ln\eta^2(T^I) + {\rm h.c.} 
\label{lth}
\eeq
where $\W^a_\alpha$ is the chiral field
strength for the factor $G_a$ of the 
gauge group.

$\L_Q$ is the one-loop quantum correction
that transforms anomalously under \ux\ and modular
transformations.  Thus, $\L_Q$ gives the field theory
anomalies canceled by the GS terms included above.
Following Refs.~\cite{gt,GNW99}, we write $\L_Q$ as
\bea
\L_Q &=& - \int d^4\theta
{E\over 8R}\sum_a \W_a^\alpha P_\chi B_a
\W_\alpha^a + {\rm h.c.},
\label{lqa} \\
B_a(L,V_X,g^I) &=& \sum_I (b-b^I_a) g^I - \dx\vx + f_a(L),
\label{lqb}
\eea
where $P_\chi$ is the chiral projection operator~\cite{1001},
$P_\chi \W^\alpha = \W^\alpha$, that reduces
in the flat space limit to $(16\Box)^{-1}\bD^2\D^2$, and
the $L$-dependent piece 
$f_a(L)$ is the ``2-loop'' contribution~\cite{gt}.
Of course, the full one-loop effective
Lagrangian has many more terms than what is shown here;
however, they are not important for our purposes.

The GS coefficients $b$ and $\dx$ appearing
in \myref{oit} must be chosen to cancel
the anomalous modular and \ux\ transformations
that the Lagrangian would have in the absence of
the GS counterterms.
It is not hard to check that the correct
choices are given by:
\beqa
\dx &=& -{1\over2\pi^2}\sum_AC^A_{a\ne X}q_A^X
= -{1\over48\pi^2} \tr Q_X
\label{dxnorm} \\
8 \pi^2 b & = & 8 \pi^2 b_a^I+ C_a 
- \sum_A(1 - 2 q^A_I)C_a^A.
\label{bdefs}
\eeqa

In the remainder of this article, 
we specialize to the case with just one 
chiral matter multiplet $\Phi$ with \ux\ 
charge $q$ and modular weights $q_I$.
Further, we take \ux\ to be the only \uone.
Then the K\"ahler potential \myref{kpt} reduces to
\beq 
K = k(L) + G + e^{G_q + 2 q V_X}|\Phi|^2, \qquad G_q = \sum_I q_I g^I,
\label{simplek}
\eeq
and \myref{oit} is unchanged.
Since the scalar component $\phi \equiv \Phi|$
must get a vev to cancel the FI term, it
is consistent to write $\phi = e^\theta$,
where $\theta$ is a complex scalar field.
Promoting this approach to a superfield
expression, we define a chiral superfield
$\Theta$ such that
\beq 
\Phi = e^\Theta. 
\eeq
We then make a field redefinition of the vector
superfield $V_X$, which is equivalent to
a \ux\ gauge transformation, 
to ``eat'' the chiral superfield $\Theta$:
\beq 
V_X \to V' = V_X + {1\over2q}\(\Theta + \bTh\),
\qquad \Phi \to \Phi' = e^{-\Theta}\Phi = 1,
\label{gaugetr}
\eeq
The field $V'$ describes a massive vector multiplet 
in the unitary gauge that we have chosen.  

To summarize,
we have made a sequence of field redefinitions
\beq
(V_X,\Phi,\bar \Phi) \to (V_X,\Theta,\bar \Theta)
\to (V').
\eeq
Because $V'$ is a {\it massive} vector multiplet,
it has more degrees of freedom than $V_X$, which
accounts for the illusion of a smaller field
content in the last step.  
Gauge invariance of $\L$
assures that we need only set $V_X \to V'$,
$\Phi \to \Phi' =1$ and $\Phibar \to \Phibar' =1$
in $\L$ to account for the
field redefinitions:
\beq
\L(V_X,\Phi,\bar \Phi) \to \L(V',1,1) .
\eeq
The expressions for $K$ and $\tL$ become
\beqa
K & = & k(L) + G + e^{G_q + 2q V'},  \label{5_2} \\
\tL & = & E \[ -3 + 2Ls(L) + L ( bG - \dx V' ) \] . 
\label{5_1}
\eeqa
Eqs.~(\ref{lqa},\ref{lth}) are given by the replacement
$\W_X^\alpha \to \W_{V'}^\alpha$, where
\beq
\W_{V'}^\alpha = - \fourth \chiproj \D^\alpha V',
\label{kja}
\eeq
and \myref{lqb} instead has $B_a(L,V',g^I)$.

In the effective theory below the scale
of \ux\ breaking we wish to
eliminate the massive $V'$ multiplet
but account for the leading effects
of its tree exchange; we accomplish
this by field redefinitions 
which eliminate linear couplings of light fields
to $V'$.  The coupling $ELV'$ which appears
in \myref{5_1} suggests we will need to
shift $L$ to accomplish this.  The presence of
$g^I$ in $K$ suggests $g^I$ might also be
involved; however, we opt to avoid redefinitions
which involve $g^I$ so that manifest modular
invariance is preserved.  An important
point in this regard is that in the supersymmetric
vacuum phase, $\evev{D_X}=0$ determines
a modular invariant vev,
\beq
\evev{e^{G_q + 2 q V_X}|\Phi|^2 |}
= \evev{e^{G_q + 2q V'}|} \not= 0.
\eeq
Thus it should be possible to have a modular invariant effective
theory after the \ux\ symmetry breaking.

We begin with a
superfield redefinition such that we have, instead of
$V'$, a massive vector superfield with vanishing
vev for its scalar component.  We do this by
re-expressing the nonvanishing vev in terms of
$L$.  With the K\"ahler potential \myref{simplek},
the FI term \myref{eq2}, rewritten in
in terms of \myref{dxnorm}, and the gauge fixing
\myref{gaugetr}, we have
\beq
D_X = \left. q e^{G_q + 2 q V'} \right| -
{ \dx \over 2 } L| .
\eeq
For the supersymmetric vacuum phase $\evev{D_X}=0$:
\beq
\bigvev{ \left. q e^{G_q + 2 q V'} \right| }
= { \dx \over 2 } \evev{L|}
\eeq
We introduce a vector
superfield $U$ with vanishing vev
(i.e., all component fields are defined
to vanish in the vacuum) such that
this condition is maintained:
\beq
q e^{G_q + 2 q V'} =  e^{2qU} {\dx \over 2} L ,
\qquad \evev{U}=0.
\eeq
This yields the redefinition
\beq
V' \equiv U + {1 \over 2q} \( \ln {\dx L \over 2q} - G_q \) .
\label{aad}
\eeq
In terms of the new
set of independent fields $(L,U,g^I)$
Eqs.~(\ref{5_2}) and (\ref{5_1}) take the form
\beqa
K & = & k(L) + G + e^{2q U} {\dx L \over 2q},  \label{5_16} \\
\tL & = & E \[ -3 + 2Ls(L) + L \( bG - \dx U
- {\dx \over 2q} \ln {\dx L \over 2q}
+ {\dx \over 2q} G_q \) \] . \label{5_17}
\eeqa

The expression for $\L_{th} + \L_Q$ is modified in two ways, when 
expressed in terms of the basis $(L,U,g^I)$.
First, the chiral field strength 
\myref{kja} now takes the form
\beqa
\W_{V'}^\alpha & = & \W_{U}^\alpha + \Y^\alpha 
\label{qe1} \\
\W_{U}^\alpha & = & - \fourth \chiproj \D^\alpha U, \label{qw4} \\
\Y^\alpha & = & - {1 \over 8q} \chiproj
\( {\D^\alpha L \over L} - \D^\alpha G_q \) .
\label{qw3}
\eeqa
Thus $\Y^\alpha$ will appear in $\L_{th} + \L_Q$.  E.g.,
\beq
\L_{th} \ni
- \int d^4\theta {E\over 8R} \sum_I b_X^I
(\W_U^\alpha + \Y^\alpha)
(\W_{U \alpha} + \Y_\alpha)
\ln \eta^2(T^I) + \hc
\label{qw2}
\eeq
Note that \myref{qe1} also effects \myref{abc}:
\beq
\chiproj L \ni - (\W \W)_{V'}
= - (\W_U + \Y)^\alpha (\W_U + \Y)_\alpha.
\eeq
Thus terms other than the chiral field
strengths of gauge multiplets appear in
the modified linearity constraints when
the redefinition \myref{aad} is made.

Second, $B_a$ in \myref{lqb}, re-expressed in terms of the new
fields, takes the form
\beq
B_a(L,V',g^I) = B_a(L,U,g^I) 
- {\dx \over2q}
\( \ln {\dx L \over 2q} - G_q \) .
\label{qw5}
\eeq
Consequently $\L_Q$ now has terms due to the shift:
\beq
\L_Q \ni \int d^4\theta
{E\over 8R} \sum_a \W_a^\alpha P_\chi
{\dx \over2q} \( \ln {\dx L \over 2q} - G_q \) 
\W_\alpha^a + {\rm h.c.}
\eeq
This is of course in addition to the $\Y^\alpha$
which appears from the $a=V'$ terms in the sum
in \myref{lqa}, analogous to \myref{qw2}.

Now consider the effect of the transformation
$(L,V',g^I) \to (L,U,g^I)$
which we have made, on
the Einstein condition.  This is a Legendre
transformation where in the new coordinates,
$(L,U,g^I)$ are to
be regarded as independent.  
In the $(L,V',g^I)$
variables the condition was satisfied:
\beq
\( {\p K \over \p L}\)_{V',g^I}
+ 2 L \( {\p S \over \p L}\)_{V',g^I} = 0,
\eeq
where from \myref{5_1}, $S(L,V',g^I)$ is given by \myref{uiw}
with $V_X$ replaced by $V'$.  However in the $(L,U,g^I)$ basis
the identification \myref{jut} now yields a different
functional (cf.~\myref{5_17}):
\beq
S(L,U,g^I) = \ts(L) + \half \( \tilde G(g^I) - \dx U \),
\label{aac}
\eeq
where for convenience we define
\beqa
\ts(L) &=& s(L) - {\dx \over 4 q} \ln {\dx L \over 2 q},
\label{qw8} \\
\tilde G(g^I) &=&
bG(g^I) + {\dx \over 2q} G_q(g^I)
= \sum_I \( b + {\dx \over 2q} q^I \) g^I.
\label{mqw}
\eeqa
We remark that the last expression defines
effective GS coefficients in the new basis.
With respect to the new variables $(L,U,g^I)$,
taking into account \myref{aad}, \myref{5_16} and \myref{aac},
the Einstein condition is no longer satisfied
due to the presence of a ``convective derivative''
term:
\beqa
\( {\p K \over \p L}\)_{U,g^I}
+ 2 L \( {\p S \over \p L}\)_{U,g^I}
&=&
\( {\p V' \over \p L}\)_{U,g^I}
\[ \( {\p K \over \p L}\)_{V',g^I}
+ 2L \( {\p S \over \p L}\)_{V',g^I} \] \nnn
&=&
{\dx \over 2q} \( e^{2qU} -1 \) = \ord{U} .
\eeqa
Further redefinitions are required, even though
this expression vanishes at vacuum according
to $\evev{U}=0$.  
We must generalize and
allow a redefinition of the linear superfield
$L$.  
However, a redefinition which involves
$L$ will generally spoil the modified linearity
conditions.  We next describe how this is avoided.

We make a transformation
$(L,U,g^I) \to (\hL,U,g^I)$ defined by
\beq
L = e^{\Dk/3} \hL
\label{6_1}
\eeq
which we associate with a Weyl transformation 
\beq
E(K) = e^{-\Dk/3} \hE, \qquad K = \hK + \Dk ,
\label{6_2}
\eeq
so as to preserve the modified linearity condition
\myref{abc}.
Here $E(K)$ denotes that $E$ is subject to
the torsion constraints which depend on $K$
whereas $\hE \equiv E(\hK)$ is subject to the
same constraints but with $K \to \hK$.
We find below that we can restrict $\Dk$ such that
\beq
\Dk = \Dk (\hL,U) 
= \alpha(\hL) U + \beta(\hL) U^2 + \ord{U^3} .
\label{6_4}
\eeq
Thus \myref{6_1} and \myref{6_4} allow us to express
$L$ as a function of $(\hL,U)$.  

From \myref{6_2}
we easily obtain $\hK$ as a function of $(\hL,U,g^I)$:
\beq
\hK = G + {\dx \hL \over 2 q } \exp \( 2qU + \third
\Dk \) - \Dk + k(L)|_{L=e^{\Dk/3} \hL} .
\eeq
We expand the last term about $\hL$ to obtain
a power series in $U$.  This will have coefficients
$k'(\hL) = dk(\hL)/d \hL$, etc.  After some
work we obtain the series
\beq
\hK = \tk(\hL) + G + \hK^U(\hL) U + \hK^{UU}(\hL) U^2 + \ord{U^3}
\eeq
with the $\hL$ dependent coefficients to
$\ord{U^2}$ given by
\beqa
\tk &=& k(\hL) + {\dx \hL \over 2 q} , \qquad
\hK^U = -{\dx \hL \over \alpha_0} \( \alpha - \alpha_0 \) ,
\label{qw7} \\
\hK^{UU} &=& \dx \hL \( q + {\alpha \over 3} - {\beta \over \alpha_0}
+ {\alpha^2 \over \alpha_0^2} \cdot {\hL \alpha_0' \over 6}
- {\alpha^2 \over 6 \alpha_0} \) ,
\label{aae}
\eeqa
where for convenience we define the quantity
\beq
\alpha_0 \equiv {3 \dx \hL \over 3 - \hL \tk'(\hL) } .
\eeq

Substitution of \myref{6_1} and \myref{6_2}
into \myref{5_17} gives $\tL$ in the new basis\footnote{The
$\hK$ which we include in the ``basis'' merely indicates
that K\"ahler covariance is now with respect to this
shifted functional.}
$(\hK;\hL,U,g^I)$.
We can write this in the form
\beqa
\tL & = &
\hE \[ -3 + 2 \hL \hS(\hL,U,g^I) \] , 
\label{yut} \\
\hS(\hL,U,g^I) & \equiv & 
{3 \over 2 \hL } \( 1 - e^{-\Dk/3} \)
+ S(L,U,g^I)|_{L=e^{\Dk/3} \hL} ,
\eeqa
where $S$ is the functional which appears in \myref{aac}.
After some manipulation $\hS$ can be brought
to the form
\beqa
\hS &=& \ts(\hL) + \tilde G(g^I)/2 + \hS^U(\hL) U
+ \hS^{UU}(\hL) U^2 + \ord{U^3}, \\
\hS^U & \equiv & {\dx \over \alpha_0} \( \alpha - \alpha_0 \), \qquad
\hS^{UU} \equiv {\dx \over 2} \( {\beta \over \alpha_0}
- {\alpha^2 \over \alpha_0^2} \cdot {\hL \alpha_0' \over 6} \).
\label{aaf}
\eeqa
To obtain these results we have made repeated
use of the identity
\beq
\tk'(\hL)+ 2 \hL \ts'(\hL) = 0 .
\label{canon2}
\eeq
which is easy to check from \myref{qw7}, 
\myref{qw8} and \myref{canon}.

We now determine the transformation
parameters $\alpha(\hL),\beta(\hL)$ such
that the Einstein condition
\beq
\( {\p \hK \over \p \hL}\)_{U,g^I}
+ 2 \hL \( {\p \hS \over \p \hL}\)_{U,g^I} = 0
\eeq
is satisfied 
and linear couplings to $U$ are eliminated.
To eliminate linear couplings
to $U$ we demand $\hK^U = \hS^U = 0$.
It is remarkable that both quantities are
proportional to $\alpha - \alpha_0$ so that we can
both eliminate the linear couplings and satisfy
the Einstein condition to $\ord{U}$ by choosing
\beq
\alpha \equiv \alpha_0 = {3 \dx \hL \over 3 - \hL \tk'(\hL) } .
\eeq
In this case the quadratic terms (\ref{aae},\ref{aaf}) simplify to
\beqa
\hK^{UU} &=& \dx \hL \( q + {\alpha \over 6} - {\beta \over \alpha}
+ {\hL \alpha' \over 6} \) , \\
\hS^{UU} &=& {\dx \over 2} \( {\beta \over \alpha}
- {\hL \alpha' \over 6} \) . 
\label{qw9}
\eeqa
From this we obtain
\beq
\hK^{UU} + 2\hL \hS^{UU} = q \dx \hL \( 1 + {\alpha \over 6 q} \)
\label{qe2}
\eeq
This in turn implies
\beq
\( {\p \hK^{UU} \over \p \hL}\)_{U,g^I}
+ 2 \hL \( {\p \hS^{UU} \over \p \hL}\)_{U,g^I} =
\dx \( q + {\alpha + \hL \alpha' \over 6} \)
- 2 \hS^{UU} .
\eeq
We demand that the RHS vanish for the Einstein
condition to be satisfied to $\ord{U^2}$.  
Using \myref{qw9} this
uniquely determines
\beq
\beta = \alpha \cdot 
\( q + {\alpha \over 6}  + {\hL \alpha' \over 3} \) .
\eeq
Finally, we use this to express the quadratic
coefficients in terms of $\alpha,\alpha'$:
\beq
\hK^{UU} = - {\dx \hL^2 \alpha' \over 6}, \qquad
\hS^{UU} = {\dx \over 2} \( q + {\alpha + \hL \alpha' \over 6} \).
\eeq

Now we must consider the effect of 
$(K;L,U,g^I) \to (\hK;\hL,U,g^I)$ on
$\L_{th} + \L_{Q}$.  First note that
$(E/R)\WaWa$ is Weyl invariant.  This leaves
the following modifications.
We replace
\beq
{E \over R} \to {\hE \over \hR}, \qquad
\W_U^\alpha \to \hat \W_U^\alpha, \qquad
\Y^\alpha \to \hat \Y^\alpha,
\eeq
where the latter two are obtained from (\ref{qw4},\ref{qw3})
by the replacement $\D \to \hat \D$,
$\bD \to \hat \bD$ to have covariance with
respect to the shifted K\"ahler potential $\hK$.
In addition the transformation \myref{6_1}
must be accounted for in (\ref{qw3},\ref{qw5}).
Because of $L$ dependence in $\Y^\alpha$,
a contribution to $\W_U^\alpha$ comes from
$\hat \Y^\alpha$ when the transformation \myref{6_1}
is made.
We find it convenient to rewrite \myref{qe1}
as a series of terms with increasing orders of $U$:
\beqa
\hat \W_{V'}^\alpha &=& H_0^\alpha + H_1^\alpha + H_2^\alpha
+ \ord{U^3} 
\label{aag} \\
H_0^\alpha & \equiv & - {1 \over 8q} \hchiproj
\( - \hat \D^\alpha G_q  + \hL^{-1} \hat \D^\alpha \hL \) 
\label{h0e} \\
H_1^\alpha & \equiv & - \fourth \hchiproj
\[ { \alpha' U \over 6q} \hat \D^\alpha \hL 
+ \( 1 + {\alpha \over 6q} \) \hat \D^\alpha U \] 
\label{qe4} \\
H_2^\alpha & \equiv & - {1 \over 24q} \hchiproj
\( \beta' U^2 \hat \D^\alpha \hL + 2 \beta U 
\hat \D^\alpha U \)
\label{h2e}
\eeqa
The second effect follows from the
reorganization of \myref{qw5}
in the new basis:
\beqa
B_a(L,V',g^I) &=& B_a(L,U,g^I) 
- {\dx \over2q} \( \ln {\dx L \over 2q} - G_q \) \nnn
&=& \hat B_a
- {\dx \over 2q} \( \ln {\dx \hL \over 2q} - G_q \)
+ U \[ {\dx \alpha \over 6q}
+ {\alpha \hL \over 3} f_a'(\hL) - \dx \] \nnn
& & + U^2 \[ {\dx \beta \over 6q} + \hL f_a'(\hL)
\( {\beta \over 3} + {\alpha^2 \over 18} \)
+ {\alpha^2 \hL^2 \over 18} f_a''(\hL) \]
+ \ord{U^3} 
\label{qe5} \\
\hat B_a & = & \sum_I (b-b_a^I) g^I + f_a(\hL)
\label{qe6}
\eeqa
Note that $\hat B_a$ is the functional which
would be present if no \ux\ anomaly existed.
Thus, the remainder of the terms are a reflection
of the effects of the anomaly.

The bosonic terms for the $U$-multiplet
coming from $\L_{th} + \L_{Q}$ are contained
in the $\theta=\bar \theta=0$
components of $\hat \D^\beta \hat \W_{V'}^\alpha 
\hat \D_\beta \hat \W_{V' \alpha}$.
By first appearances, it is a nontrivial task
to extract the leading terms from 
Eqs.~\myref{aag}-\myref{h2e}.  However, we now
show that in the supersymmetric vacuum where
auxilliary fields have vanishing vevs and $\evev{U}=0$,
significant simplifications occur.  For the
purpose of illustration we extract 
$\hat \D^\beta  \hat \W_{U}^\alpha 
\hat \D_\beta  \hat \W_{U \alpha}$.

From \myref{h0e}-\myref{h2e} it is not hard to
show that
\beqa
\Dbu H_0^\alpha &=& - {1 \over 8q \hL} 
\Dbu \hchiproj \Dau \hL + \cdots \\
\Dbu H_1^\alpha &=&
\(1 + {\alpha \over 6q} \) \Dbu \hat \W_U^\alpha
- {\alpha' U \over 24 q} \Dbu \hchiproj \Dau \hL + \cdots 
\label{cah} \\
\Dbu H_2^\alpha &=&
{\beta U  \over 3q} \Dbu \hat \W_U^\alpha
- {\beta' U^2 \over 24 q} \Dbu \hchiproj \Dau \hL + \cdots 
\eeqa
where $\cdots$ indicates terms containing only
$\Dad \hat \bD_{\dot \alpha} \hL$, $\Dad \hat \bD_{\dot \alpha} U$,
$\Dad \Dbd \hL$, $\Dad \Dbd U$ and hermitian conjugates
of these.  Such terms do not contribute to the
$U$-multiplet vector boson field strength or auxilliary
field $D_U$ contained in $\hat \D^\beta  \hat \W_{U}^\alpha$,
so they are irrelevant for our immediate purpose.

It can be shown that
\beqa
\Dbu \hchiproj \Dau \hL &=&
\epsilon^{\alpha \beta} \Lambda(\hL,U) \hat \D^\gamma 
\hat \W_U^\delta \hat \D_\gamma \hat \W_{U \delta} , \nnn
\Lambda(\hL,U) & \equiv & \(1 + {\alpha \over 6q} \)^2
\( 1 + {8 \hL \hK' \over 3 - \hL \hK'} \) + \cdots .
\eeqa
It follows that
\beqa
&& \Dbu \hat \W_{V'}^\alpha 
\Dbd \hat \W_{V' \alpha}  =  \nnn
&& \qquad \[ 2 \sum_{i=0}^2 \evev{\hat \D^\gamma H_{i \gamma}}
\Lambda(\hL,U) \( {-1 \over 8q} \)
\( \hL^{-1} + {\alpha' U + \beta' U^2 \over 3q} \)
+ \( 1 + { \alpha \over 6q} + { \beta U \over 3 q} \)^2
\] \nnn
&& \hspace{1in} \times \Dbu \hat \W_{U}^\alpha 
\Dbd \hat \W_{U \alpha} + \ord{U^4} + \cdots .
\label{caf}
\eeqa
Here we have kept explicit several terms which are
$\ord{U^3}$, so that we may illustrate terms which
drop out at leading order due to $\evev{U}=0$.

Examination of $\Dbu H_{i}^\alpha$ shows that
$\evev{\hat \D^\gamma H_{i \gamma}}=0$ for each $i=0,1,2$,
provided supersymmetry is unbroken.  Then \myref{caf}
reduces to
\beq
 \Dbu \hat \W_{V'}^\alpha 
 \Dbd \hat \W_{V' \alpha} \ni 
\( 1 + {\alpha \over 6q}\)^2
( \Dbu \hat \W_{U}^\alpha 
 \Dbd \hat \W_{U \alpha}) .
\eeq
Thus it can be seen that we need only look at
$\Dbu H_1^\alpha \Dbd H_{1 \alpha}$ and that
furthermore we only need the first term in \myref{cah}.
Similar arguments can be made for the other leading
terms relevant to the $U$-multiplet
coming from $\Dbu \hat  \W_{V'}^\alpha 
\Dbd \hat  \W_{V' \alpha}$.

From \myref{qe4} it is easy to show that
\beq
H_1^\alpha = \( 1 + {\alpha \over 6q} \) \hat \W_U^\alpha 
- \fourth \hchiproj \( {\alpha' U \over 6q} \Dau \hL \)
- \( { 1 \over 24 q} \) \( \hat \bD^2 \alpha \Dau U
+ 2 \hat \bD_\alphadot \alpha \hat \bD^{\alphadot} \Dau U \) .
\label{qe3}
\eeq
We note that the coefficient
of $\hat \W_U^\alpha$ is proportional to \myref{qe2}.
Thus we extract the leading term to obtain
\beqa
\L_{th}^{(\hat \W \hat \W)_U} & = &
-\superint \,{\hat E \over 8 \hat R}
\sum_I b_X^I \( 1 + {\alpha \over 6q} \)^2
(\hat \W \hat \W)_U \ln \eta^2(T^I) + {\rm h.c.} \\
\L_{Q}^{(\hat \W \hat \W)_U} & = &
-\superint \,{\hat E \over 8 \hat R}
\( 1 + {\alpha \over 6q} \)^2
\hat \W_U^\alpha P_\chi \[ \hat B_a 
- {\dx \over 2q} \( \ln {\dx \hL \over 2q} - G_q \) \]
\hat \W_{U \alpha} + {\rm h.c.},
\eeqa

We note that linear interactions
with $U$ arise from the terms
\beq
\hat \D^\beta \hat \W_{V'}^\alpha 
\hat \D_\beta \hat \W_{V' \alpha}
\ni
2 \hat \D^\beta H_0^\alpha 
\hat \D_\beta H_{1 \alpha} 
\eeq
Detailed examination shows that these terms are
higher dimensional and either involve derivatives
or auxiliary fields; thus they are suppressed
and we neglect them in our leading order analysis.
When shifts due to supersymmetry breaking are
studied, it will be necessary to account for
the effects of terms other than \myref{qe3}.

For the nonanomalous factors $G_a$ of the gauge
group, the chiral field strengths are merely replaced according to
$\W_a^\alpha \to \hat \W_a^\alpha$ under the redefinitions
\myref{6_1} and \myref{6_2}.
The shift in \myref{qe5} yields
\beq
\L_Q^{a \not=X} = \hat \L_Q^{a \not=X}
+ \Delta \L_Q^{a \not=X}
\eeq
where $\hat \L_Q^{a \not=X}$ is the quantum
correction in the absence of a \ux\ (i.e., with
$B_a$ replaced by the $\hat B_a$ 
of \myref{qe6} in \myref{lqa}) and
\beqa
\Delta \L_Q^{a \not=X} & = &
-\superint \,{\hat E \over 8 \hat R}
\hat \W_a^\alpha P_\chi \[ B_a(L,V',g^I) - \hat B_a \]
\hat \W_{a \alpha} + {\rm h.c.} \nnn
& = &
{\dx \over 2q} \superint \, {\hat E \over 8 \hat R}
\hat \W_a^\alpha P_\chi \( \ln {\dx \hL \over 2q} - G_q \)
\hat \W_{a \alpha} + \ord{U} + \hc 
\eeqa
The shift $(\dx / 2q) P_\chi \ln (\dx \hL / 2q)$
cancels the contribution to the Yang-Mills kinetic terms 
arising from $\ts -s$ in \myref{qw8},
and restores the gauge coupling to its original form.
The shift $(\dx / 2q) P_\chi G_q$ cancels a
corresponding shift in $\int \tL$, so as to
maintain modular invariance.

A careful examination of the component expansion of the
redefined superfield Lagrangian described above yields for
the mass of the $U$ vector multiplet
\beq m^2 = \bigvev{ {1\over s(\hat \ell)}
\(1 + {\alpha(\hat \ell) \over 6q} \)^{-2}
\[\hK^{UU}(\hat \ell) + 2 \hat \ell \hS^{UU}(\hat \ell)\]} =
\bigvev{ {q \dx \hat \ell \over s(\hat \ell)
[1 + \alpha(\hat \ell)/6q]}}
\label{vmass}
\eeq

Next we perform the component field 
calculation.  We will not assume WZ gauge.  This is
not as daunting as it may seem, 
because we can use results from \cite{BGWa}. To 
simplify matters we neglect $T$-moduli here.  
First consider a general
vector superfield in global supersymmetry:
\begin{eqnarray}
V &=& C + i \theta \chi - i \bar{\theta} \bar{\chi} +
\theta^2 h + \bar{\theta}^2 \h + \theta \sigma^m 
\bar{\theta}  a_m \nonumber\\
& & + i\theta^2 \bar{\theta}(\bl + {i \over 2} \bar{\sigma}^m 
\partial_m\chi) -i\bar{\theta}^2 \theta ( \lambda + {i \over 2}
\sigma^m \partial_m \bar{\chi}) \nonumber \\
& & + {1\over2}\theta^2 \bar{\theta}^2
(D + {1 \over 2} \Box C). \label{expansion}\end{eqnarray} 
In the WZ gauge  $C = \chi = h = 0$.  We get a massive vector field
when a massless one eats a chiral multiplet $\Theta = (\theta,\chi,h)$
with $C= \theta + \bth$ in U-gauge.  So in supergravity it is natural to 
define 
\beq H = - {1\over4}\DaDa V, \quad \H = - {1\over4}\DbDb V, \quad
h = \l H\r, \quad \h = \l\H\r. \label{defh}\eeq
Then comparing with Eqs.~(3.4)-(3.6) of \cite{BGWa} we have 
\bea \l\frac{1}{2}[\,\D_{\alpha},{\bar \D}_{\dot{\alpha}}\,]V\r &=&
\sigma^{m}_{\alpha\dot{\alpha}}a_{m} - 
\frac{2}{3}C\sigma^{a}_{\alpha\dot{\alpha}} b_{a} =
\sigma^{m}_{\alpha\dot{\alpha}}v_{m},\quad C =\l V\r,\nonumber\\
\l-(\bar{\D}^{2}-8R)V\r &=& 4\h - {4\over3}MC,\quad
\l-(\D^{2}-8\R)V\r = 4h - {4\over3}\M C,
\label{vcomps} \eea
where 
\beq -\,\frac{1}{6}M\,=\l R\r,\quad -\,\frac{1}{6}\bar{M}\,=\l\R\r,\quad
-\,\frac{1}{3}b_{a}\,=\l G_{a}\r,\label{gravaux}\eeq
are the auxiliary components of supergravity multiplet. In
addition
\bea - 4F &\equiv& - \l\DaDa\(\DbDb - 8R\)V\r = - \l\DaDa\DbDb V\r 
+ 8C\l\DaDa R\r + {16\over3}Mh, \nonumber \\
- 4\bF &\equiv& - \l\DbDb\(\DaDa - 8\R\)V\r = 
- \l\DbDb\DaDa V\r + 8C\l\DbDb\R\r + {16\over3}\M\h.\label{vaux}\eea
 Further comparison with~\cite{BGWa} gives
\bea F-\bar{F} &=& 
4\(i\nabla^{m}v_{m} + {2i\over3}C\nabla^{m}b_{m} + \h\bar{M}- hM\), 
\nonumber \\
\l\DaDa R\r - \l\DbDb\R\r &=& 4i\l\D^aG_a\r = - {4i\over3}\nabla^mb_m,
\nonumber \\ \l\DbDb\DaDa V\r - \l\DaDa\DbDb V\r &=&  
- 16i\nabla^{m}v_{m} - {32\over3}\(\M\h - Mh + ib_m\pp^m C\).
\label{diffs}\eea
The expression for $F+\bar{F}$ contains the auxiliary 
field $D$:
\bea D &=&\l\frac{1}{8}\D^{\beta}(\bar{\D}^{2}-8R)\D_{\beta}V\r 
   =\l\frac{1}{8}\D_{\dot{\beta}}(\D^{2}-8\R)\D^{\dot{\beta}}V\r.\eea  
We can evaluate $D$ using Eqs. (3.25) and (3.28) of
\cite{BGWa}. We drop all superfields
except $V$ and make the substitutions 
$$k(V) = \ln(V) + g(V) \to V, \quad V g_1 + 1 = V k'(V) \to V, 
\quad V^2g_2 - 1 =
V^2k''(V) \to 0, $$ $$ X_\alpha \to -{1\over8}\(\DbDb - 8R\)\D_\alpha V, \quad
\l \D^\alpha X_\alpha\r \to - D. $$  
Then we obtain
\bea 2D &=& {1\over8}\(\l\DbDb\DaDa V\r + \l\DaDa\DbDb V\r\) 
+ {4\over3}\(v^mb_m - hM - \h\M\) - 2\Box C.\label{2d}\eea
So finally we get 
\beq {1\over8}\l\DaDa\DbDb V\r = D + {4\over3}\h\M - {2\over3}v^mb_m + 
\Box C + i\nabla^{m}v_{m} + {2i\over3}b_m\pp^m C.\label{d4v}\eeq
The $V$-dependent part of the (bosonic) Lagrangian is (aside from the usual gauge term)
\beq \L(V) = e\[s(V) + \s(V) - r(V)\M - \br(V)M - \l{1\over2}\D^\alpha X_\alpha(V)\r\],
\label{lv0} \eeq
with (dropping fermions)
\bea r(V) &=& {\dx\over8}\l\(\DbDb - 8R\)LV\r = - {\dx\over2}\h\ell, 
\nonumber\\
 s(V) &=& - {\dx\over8}C\(F^2 - iF\tF - 2D^2\) 
+ {\dx\over4}\(v^m -i\pp^mC\)B_m \nonumber \\ & & 
- {\dx\ell\over4}D - {\dx\ell\over2}\h\M
+ {\rm total \;deriv.}, \nonumber \eea
\bea - {1\over2}\D^\alpha X_\alpha(V) &=& 
{1\over16}\Da\(\DbDb - 8R\)\Dc e^{2qV}|\Phi|^2
= e^{2qC}\bigg\{qD|\phi^2| + |F_\Phi + 2qh\phi|^2 \nonumber \\ & &
- \[\pp^m\phi + q\phi\(iv^m +\pp^mC\)\]\[\pp_m\ph - q\ph\(iv_m
- \pp_mC\)\]\bigg\},\label{lv}\eea
where the one-form $B_m$ is dual to a linear combination of the curl
of a two-form $b_{mn}$ and the Yang-Mills Cern-Simons form 
$\omega^{YM}_{lmn}$.  The first term on the RHS of $s(V)$ 
is canceled by the quantum correction. All the terms linear in $\M,b_m$
cancel in $\L(V)$, so in the absence of a superpotential, they vanish
by their equations of motion.  However we wish to keep local supersymmetry
explicit down to scales where supersymmetry is broken; hence we will
not set these auxiliary fields to zero.  Now set 
\beq \phi = \sigma e^{i\alpha}, \quad \pp\phi = \(\pp\sigma + i\sigma
\pp\alpha\) e^{i\alpha}.\label{phiredef}\eeq
Including the standard Yang-Mills term, the Lagrangian for $D$ is
\beq \L(D,C) = {1\over2g_s^2}D^2 + \(e^{2qC}q\sigma^2 - {\dx\ell\over2}\)D.
\label{ld}\eeq
The equation of motion for $D$ gives
\beq D = - {g_s^2}\(e^{2qC}q\sigma^2  -
{\dx\ell\over2}\) = - g_s^2{\dx\ell\over2}\(e^{2qc} - 1\) =
- g_s^2\dx\ell qc + O(c^2). \label{deom}\eeq
The field redefinitions in (\ref{deom}) are the scalar projections of
(\ref{gaugetr}) and (\ref{aad}): 
\beq C' = C + {1\over q}\ln\sigma,\quad  e^{2qC'} =
{\dx\ell\over2q}e^{2qc},
\quad \lvev c\rvev=\lvev\l U\r\rvev = 0.\label{scalarredefs}\eeq
The one-form plus axion Lagrangian is
\bea \L(B,v,b,\alpha) &=&  \frac{k'(\ell)}{4\ell}B^{m}\!B_{m}
+ {\dx\over2}v^mB_m - \sigma^2e^{2qC}\(qv^m + \pp^m\alpha\)
\(qv_m + \pp_m\alpha\) -{1\over9}\(\ell k' - 3\)b_mb^m
\nonumber \\ &=& \frac{k'(\ell)}{4\ell}B^{m}\!B_{m} +
 {\dx\over2}v'^mB_m - q^2{\dx\ell\over2}v'^mv'_m 
- {1\over9}\(\ell k' - 3\)b_mb^m + O(c), 
\nonumber \\ v'_m &=& v_m + {1\over q}\pp_m \alpha.\label{1form}\eea
 We dropped from (\ref{1form}) a term (dropping also a total derivative)
$$ {\dx\over2q}\pp^m\alpha B_m = - {\dx\over2q}\alpha\nabla^mB_m =
-{\dx\over4q}\alpha F\tF$$
that is canceled by a shift in $\L_Q$ under the redefinition
 $v\to v'$, which
is just an ordinary gauge transformation,
{\it i.e.} the vector projection of (\ref{gaugetr}).
The vector projection of the first equality in (\ref{aad}) is:
\beq v'_m = u_m + {1\over2q\ell}\(B_m - {2\ell\over3}b_m\).\label{uprime}\eeq
This gives
\bea \L(B,b,v') &=& \frac{k'(\ell)}{4\ell}B^{m}\!B_{m}
-{1\over9}\(\ell k' - 3\)b_mb^m
+ {\dx\over2}v'^mB_m - {q\dx\ell\over2}v'^mv'_m + O(c), \nonumber \\
&=& \frac{\tk'(\ell)}{4\ell}B^{m}\!B_{m}
-{1\over9}\(\ell \tk' - 3\)b_mb^m
+ {\dx\ell\over3}u^mb_m - {q\dx\ell\over2}u^mu_m
+ O(c)\label{l1form}\eea
 Solving for $b_m$ then gives
\bea \L(B,u) &=& \frac{\tk'(\ell)}{4\ell}B^{m}\!B_{m}
- {q\dx\ell\over2}\(1 + {\alpha\over6q}\)u^mu_m
+ O(c),\nonumber\\
b_m &=& {\dx\ell\over2}{3u_m\over\ell\tk' - 3},\label{beom}\eea
which gives the same mass as in (\ref{vmass}) 
when we take into account the
normalization of the kinetic term for $u_m$. Indeed, when we
substitute the last equality in \myref{beom} into (\ref{uprime}), we
see $u_m$ gets renormalized relative to $v'_m$ by a factor 
$(1 + \alpha/6q)$, and hence its kinetic term is multiplied
by a factor $(1 + \alpha/6q)^2.$  Alternatively, we can redefine $b_m$:
\bea\L(B,\hb,u) &=& \frac{\tk'(\ell)}{4\ell}B^{m}\!B_{m}
-{1\over9}\(\ell\tk' - 3\)\hb_m\hb^m - {q\dx\ell\over2} 
\(1 + {\alpha\over6q}\)u^mu_m
+ O(c),\nonumber \\ & & 
\quad \hb_m = b_m - {3\dx\ell u_m\over2(\ell\tk' - 3)}.
\label{alt1form}\eea 
The last equality in \myref{alt1form}, which again provides the
correct normalization of the $u_m$ kinetic 
term,\footnote{After these field redefinitions the 
squared field strength $v_{mn}v^{mn}$ also contains higher 
derivative terms of the type that we have neglected throughout.}
is the vector projection of (\ref{6_1})
with $\hat B = B$, up to order $c$ corrections.  (Note that $\hat B = B$
preserves the relation $\nabla^mB_m = F\tF/2$ that follows from 
\myref{abc}.) The equations of motion for $F_\Phi,h,\h$ give 
\beq F_\Phi + 2qh\phi = 0.\label{fheom}\eeq
For the $\D^2$ projections of (\ref{gaugetr}) and (\ref{aad}) we have
\beq h' = - {1\over4}\D^2V' = h + {1\over2q\phi}F_\Phi 
= f + {1\over6q}\M,\quad f \equiv - {1\over4}\D^2U,\label{auxredefs} \eeq   
so in (\ref{lv}) $e^{2qC}|F_\phi + 2qh\phi|^2\to 2q\dx\ell e^{2qc}|f 
+ {1\over6q}\M|^2$
which still vanishes when the equations of motion for $f$ are imposed.  
However if we keep the
full $M$-dependence, we can cast these terms 
in a form that is the supersymmetric 
counterpart of the one-form Lagrangian:
\bea \L(M,f) &=& {1\over9}(\ell k' - 3)M\M 
+ 2q\dx\ell e^{2qc}|f + {1\over6q}\M|^2 \nonumber \\
&=& {1\over9}(\ell\tk' - 3)M\M + 
{\dx\ell\over3}\(\f\M + fM\) + 2\dx\ell|f|^2 + O(c)
 \nonumber \\
&=&{1\over9}(\ell\tk' - 3)\hM\hbM 
+ 2q\dx\ell\(1 + {\alpha\over6q}\)|f|^2 
+ O(c) = \L(\hM,f),\nonumber \\ \hM &=& M + {3\dx\ell\f\over\ell\tk'-3}.
\label{auxredef}\eea
The last equality is the $D^2$ projection of (\ref{6_1}), and
$\L(\hM,f)$ is the supersymmetric counterpart of $\L(B,\hat b,u)$
in (\ref{alt1form}).\footnote{In the expansion 
(\ref{expansion}) the field $h$ is
related to Wess-Bagger auxiliary fields by $h = -{i\over2}\(N-iM\)$,
which explains the extra factor 4 in the $|f^2|$ term,
relative to the $u^2$ term in (\ref{alt1form}); {\it cf.}~Eq.~(6.18) of
\cite{WB92}.)}
Since $D$ is invariant under the gauge transformation (\ref{gaugetr}), we have
\bea D' &=& D = D_U + \l\frac{1}{8}\D^{\beta}(\bar{\D}^{2}-8R)
      {1\over2q}\D_{\beta}\ln\({\dx L\over2q}\)\r \nonumber\\
&=& D_U + {1\over2q(3 - \ell k')}(\dx\ell D + 3D^2) + X + O(cD), \nonumber \\
D &=& \(1 + \alpha/6q\)\(D_U + X\) + O(D^2),
\nonumber \\ X &=& - {3\ell^{-1}\Box\ell + r\over2q(3 - \ell k')}
+ \cdots, \label{stuff}
\eea
where the ellipses represent terms quadratic in auxiliary
fields and/or derivatives of fields.  
Since 
$$
{\pp\L\over\pp D_U} = {\pp D\over\pp D_U}{\pp\L\over\pp D},
$$
is solved by ${\pp\L/\pp D} = 0$, 
the  scalar Lagrangian should not be modified by this redefinition.  However
in order to cast the full component Lagrangian in a manifestly 
supersymmetric form, we do not eliminate the auxiliary field $D$
 or $D_U$. Expressing \myref{ld} in terms of $D_U$, we have
\beq\L(D,C) = \L(D_U,c) + \L(X,c) + \L(X,D_U),\eeq
where 
\beq \L(D_U,c) = {1\over2}g_X^{-2}D_U^2 
+ \(1 + {\alpha\over6q}\)qc\dx\ell D_U + O(c^3),  \quad 
 g^{-1}_X = g^{-1}_s\(1 + {\alpha\over6q}\), \label{ld2}\eeq
contains the supersymmetric counterparts of the kinetic term
for $u_m$, with $g_X$ the effective \ux\, gauge coupling constant,
and of the last terms in \myref{alt1form} and \myref{auxredef}.
To evaluate the terms containing $X$, we first write
\beq \L(X,c) = - \(1 + {\alpha\over6q}\)qc\dx\ell X = -{\alpha\over2}c
\(\ell^{-1}\Box\ell + {1\over3}r + \cdots\).\label{lxc}\eeq
Evaluating the first term by partial integration, and performing a
Weyl transformation 
\beq  -{1\over2}\sqrt{g}r\(1 + {\alpha c\over3}\) = -{1\over2}\sqrt{\hat g}\[\hat r
+ {\alpha^2\over6}\nabla^m c\nabla_m c - \Box(\alpha c)\] + c(\cdots) \eeq
to restore the Einstein term to canonical form, we obtain from \myref{lxc}
a contribution 
\beq \del\L = {\alpha\over2\ell}\nabla^m\ell\nabla_m c -
{\alpha^2\over12}\nabla^m c\nabla_m c + c(\cdots),\eeq
where the ellipses have the same meaning as in \myref{stuff}.
Combining this with  
\bea \L_{KE}(\ell,\sigma,C) &=& -\,\frac{k'(\ell)}{4\ell}
\nabla^{m}\!\ell\,\nabla_{\!m}\!\ell -
\nabla^m(e^{qC}\sigma)\nabla_m(e^{qC}\sigma)\nonumber \\ 
&=& -\[\frac{\tk'(\ell)}{4\ell}
\nabla^{m}\!\ell\,\nabla_{\!m}\!\ell + {\dx\over2}\nabla^m\ell\nabla_m c +
{q\dx\ell\over2}\nabla^m c\nabla_m c\]\[1 + O(c)\], \label{ske}\eea
the full scalar kinetic energy term take the form
\bea  \L_{KE}(\ell,c) &=& -\[\frac{\tk'(\ell)}{4\ell}\(
\nabla^{m}\!\ell\,\nabla_{\!m}\!\ell - 2{\dx\ell^2\over3-\ell\tk'}
\nabla^m\ell\nabla_m c\) +
\({q\dx\ell\over2} + {\alpha^2\over12}\)\nabla^m c\nabla_m c\]\[1 + O(c)\]
\nonumber \\ &=& -\[\frac{\tk'(\hel)}{4\hel}
\nabla^{m}\!\hel\,\nabla_{\!m}\!\hel -
{\dx\hel q\over2}\(1 + {\alpha\over6q}\) c\Box c\]\[1 + O(c)\],
\nonumber \\ \ell &=& \hel\[1 + \alpha c/3 + O(c^2)\].\label{rhodef}\eea
The last equality is the scalar projection of \myref{6_1}, and
the Lagrangian in \myref{rhodef} contains the supersymmetric
counterparts of the terms quartic in $B_m$ and $v_m$,  respectively.
Now in terms of the hatted variables, the $\Box c$ terms cancel in $X$,
and we have
\bea X &=& - {3\ell^{-1}\Box\ell + r\over2q(3 - \ell k')}
+ \cdots = - {3\hel^{-1}\Box\hel + \hat r\over2q(3 - \hel k')}
\[1 + O(c)\] + \cdots\nonumber \\ &=& 
- \({3\over\tk'}{\pp V\over\pp\hel} - 4V\){1\over2q(3 - \hel k')}
\[1 + O(c)\] + \cdots = O(c^2) + \cdots, \eea
where in the second line we used the equations of motion 
for the metric $g_{\mu\nu}$ and the scalar $\hel$.
Then finally we obtain 
\beq\L(X,D_U) =  {1\over2}g_X^{-2}\(2D_UX + X^2\) =
 O(c^3).\eeq
Now solving \myref{ld2} for $D_U$ to obtain the scalar potential,
and using the normalization of the $c$ kinetic term in \myref{rhodef},
we again recover the mass \myref{vmass} for $c$.

Note that if we had first solved for $D$ as in \myref{deom},
the scalar Lagrangian would have taken the form
\bea \L(\ell,c) &=& - {1\over2}g^2\dx^2{\tilde\ell}^2q^2c^2
 -\[\frac{\tk'(\tilde\ell)}{4\tilde\ell}
\nabla^{m}\!\tilde\ell\,\nabla_{\!m}\!\tilde\ell +
{\dx\tilde\ell\over4}\(2q-{\dx\over\tk'}\)\nabla^m c\nabla_m c\]\[1 + O(c)\]
+ O(c^3),\nonumber \\ 
\ell &=& \tilde\ell - {\dx\tilde\ell\over\tk'(\tilde\ell)}c.
\label{ske2}\eea
This would give 
\beq m^2_c =
{2g_s^2q^2\dx\tilde\ell\tk'\over2q\tk'-\dx} = g_s^2\(q\dx\tilde\ell +
{\dx^2q\tilde\ell\over2q\tk' - \dx}\) = g_s^2\(q\dx\tilde\ell +
{\dx^2\tilde\ell\over2k'}\)
\eeq
for the mass of the scalar component $c$, in agreement with the mass
of the \ux\, vector boson found in~\cite{bdp} using the chiral multiplet
formulation for the dilaton.  The same result can be obtained in the
linear multiplet formulation by first eliminating
$b_m$ by its equation of motion, and then eliminating the $B_mv^m$
coupling by a redefinition of $B_m$, although this breaks
both the linearity condition and manifest supersymmetry in the
two-form and dilaton kinetic terms. Thus the result for the masses
is prescription-dependent, which suggests that they are not really
physical,  {\it i.e.} that they do not correspond to poles in 
the propagators. What we have shown here is that it is possible to
maintain explicitly local supersymmetry and the linearity condition
(as well as modular invariance when the moduli are included)
by making consistent superfield redefinitions.

The work performed in this article suggests further
research, which is in progress \cite{GG02}:  first,
the incorporation of complicating aspects:
systems with more matter fields; dynamical supersymmetry
breaking by gaugino condensation, and  second,
the extraction of supersymmetry breaking soft
parameters and the consequent electroweak scale
phenomenology.

Already without these more realistic features, we have arrived at some
interesting conclusions.  The modified linearity constraints
\myref{abc} are significantly modified when rewritten in the light
field basis, due to the several terms in \myref{aag}.  Yet the
disturbance is in some sense minimal since the new pieces are
compactly encoded (at a superfield level, no less) in the functional
\myref{aag}.  In addition, the one-loop effective contribution is
modified significantly in the new basis, as can be seen in
\myref{qe5}.  We have shown how to fix to unitary gauge at the
superfield level.  We have eliminated the important linear couplings
to the heavy vector multiplet.  The remaining linear couplings appear
through \myref{aag}, but are all higher-order derivative interactions
or suppressed by the supersymmetry breaking scale due to the presence
of auxiliary fields.  We do not seem to be able to dispose of these
terms in a simple fashion.  Modular invariant field redefinitions were
made all along.  Because of this, the effective theory of light fields
is manifestly modular invariant with modified modular weights
for \ux-charged chiral multiplets.  We have also shown that our
results at the superfield level can be reproduced at the component
field level.

\vspace{0.20in}

\noindent {\bf \Large Acknowledgments}

\vspace{5pt}

\noindent 
We wish to thank Pierre
Bin\'etruy for disucssions during the early stages of this work. This
work was supported in part by the Director, Office of Science, Office
of High Energy and Nuclear Physics, Division of High Energy Physics of
the U.S. Department of Energy under Contract DE-AC03-76SF00098 and in
part by the National Science Foundation under grant PHY-0098840.

\end{document}